\documentclass[a4paper,12pt]{article}

\usepackage[utf8]{inputenc} 
\usepackage[francais]{babel}
\usepackage[OT1]{fontenc}       
\usepackage{graphicx}
\usepackage{jdf}  
\usepackage{times}
\usepackage{url}

\begin{document}

\renewcommand{\figurename}{Figure~}
\renewcommand{\tablename}{Tableau~}
\date{}

\title{ On Using Agent-based Modeling and Simulation\\for Studying Blockchain Systems
 \\[1.3em]
 Sur l'Utilisation de la Modélisation et de la Simulation basées Agents pour Etudier les Systèmes de Chaînes de Blocs
}

\author{\small
        \"{O}nder G\"{u}rcan\footnote{ORCID: 0000-0001-6982-5658} \\
        ~\\
        \small{Université Paris-Saclay, CEA, List}
        \\
        \small{F-91120, Palaiseau, France}\\
        \small{onder.gurcan@cea.fr}
}

\parskip 3mm
\maketitle

\thispagestyle{empty}


Bitcoin \cite{Nakamoto2008} is the core of decentralized crypto-currency systems.
The underlying data structure of Bitcoin is called the blockchain in which transactions of digital coins between accounts are batched in so-called blocks, where each block is appended to the last one in a cryptographic way to make the malicious/accidental change of blocks content very hard.
%
Participants following this protocol can create together a distributed, economical, social and technical system where anyone can join/leave 
and perform transactions in-between without neither needing to trust each other nor having a trusted third party.
It is a very attractive technology since it maintains a \textit{public}, \textit{immutable} and \textit{ordered} log of transactions which guarantees an \textit{auditable} ledger accessible by anyone. 

Technically speaking, a blockchain system is an open and distributed transactional system composed of participants called \textit{users} and \textit{block creators}.
All participants of a blockchain system store unconfirmed transactions in their memory pools and confirmed transactions in their blockchains.
However, it should be noted that, a participant may enter into several blockchain systems at the same time and can have different objectives and expectations in each. 


Moreover, blockchain systems have the following three characteristics.

\begin{itemize}
\item \textit{Blockchain systems are distributed systems.} As Leslie Lamport says\footnote{Leslie Lamport, \textit{Distribution}, \url{https://www.microsoft.com/en-us/research/publication/distribution/}, E-mail message sent to a DEC SRC bulletin board at 12:23:29 PDT on 28 May 1987.}, a distributed system is one in which the \textit{failure} of a computer you did not even know existed can render your own computer unusable.
Most distributed systems (including blockchain systems) are designed with \textit{fault tolerance} as one of the main objectives in order to achieve high availability and storing or processing distributed data.
\item \textit{Blockchain systems are social organizations.}
A social organization can be defined as formal or informal groups of interrelated individuals (agents) who pursue a \textit{collective goal} and who are embedded into an \textit{environment} \cite{Ostrom2009}.
Moreover, the blockchain (data structure) is a physical manifestation of the interactions of users.
Blockchain systems facilitates \textit{cooperation} by getting self-interested, distrustful people to work together, even when narrow self-interest would seem to dictate that no individual should take part.
Blockchain systems have highly volatile dynamics, conflict of individual/collective goals (e.g., users want lower fees while block creators want higher fees) and continuous enter/exit dynamics \cite{Gurcan2017}.
\item \textit{Blockchain systems are obviously economical systems.}
An economical system, as any other complex system, reflects a \textit{dynamic interaction} of a large number of different agents, not just a few key agents.
The resulting systemic behavior, observable on the \textit{aggregate level}, often shows consequences that are hard to predict (e.g., the transaction fees)  which cannot be simply explained by the behaviors of a few major agents.

\end{itemize}

Furthermore, it is a very active and dynamic ecosystem where new blockchain platforms and algorithms are developed continuously, due to the interest of industries to the technology.

Thus, there is a need for a simulation framework, which is develop as a \textit{software} using modern engineering approaches (e.g., modularity - i.e. model reuse-, testing, continuous development and continuous integration, automated management of builds, dependencies and documentation) and agile principles, (1) to make rapid prototyping of industrial cases and (2) to carry out their feasibility analysis in a realistic manner (i.e., to test hypothesis by simulating complex experiments involving large numbers of participants of different types acting in one or several blockchain systems).

Besides, we anticipate a challenging and interdisciplinary
research agenda in blockchain systems, built upon a
methodology that strives to capture the rich process resulting
from the interplay between the behavior of agents and the
dynamic interactions among them. To be effective, however,
simulation studies providing insights into blockchain systems
from massive data analysis, theory encompassing the
appropriate description of agents and their interactions, and
a systemic perspective bestowing a new understanding of
global effects as coming from varying network interactions
are needed. We predict that such studies will create a
more unified field of blockchain systems that advances our
understanding and leads to further insight. However, such
studies are not trivial and very hard to conduct without
dedicated tool support.

We claim that such a simulation framework needs to be a deployed \textit{agent-based modeling and simulation} platform which allows to recreate the dynamics of systems at an agent level, so that the impact of actions executed by an algorithm under different scenarios can be tested and analyzed in granular detail.
We also claim that to better capture and model the requirements of
blockchain systems, it should follow an organization centered approach \cite{Ferber2003}
rather than an agent centered one.
To the best of our knowledge, there is no such platform yet.
We are currently active working on this and our first results \cite{Lagaillardie2019} are promising.

\bibliographystyle{ieeetr}
\bibliography{references}

\begin{thebibliography}{1}

\bibitem{Nakamoto2008}
S.~Nakamoto, ``Bitcoin: A peer-to-peer electronic cash system,'' 2008.
\newblock \url{https://bitcoin.org/bitcoin.pdf}.

\bibitem{Ostrom2009}
E.~Ostrom, ``A general framework for analyzing sustainability of
  social-ecological systems,'' {\em Science}, vol.~325, no.~5939, pp.~419--422,
  2009.

\bibitem{Gurcan2017}
{\"O}.~G{\"u}rcan, A.~Del~Pozzo, and S.~Tucci-Piergiovanni, ``On the bitcoin
  limitations to deliver fairness to users,'' in {\em On the Move to Meaningful
  Internet Systems. OTM 2017 Conferences} (H.~Panetto, C.~Debruyne, W.~Gaaloul,
  M.~Papazoglou, A.~Paschke, C.~A. Ardagna, and R.~Meersman, eds.), (Cham),
  pp.~589--606, Springer Int. Publishing, 2017.

\bibitem{Ferber2003}
J.~Ferber, O.~Gutknecht, and F.~Michel, ``{Agent/Group/Roles: Simulating with
  Organizations},'' in {\em {ABS'03: Agent Based Simulation}} (M.~J.P., ed.),
  (Montpellier (France)), p.~12, Apr. 2003.
\newblock April 28-30.

\bibitem{Lagaillardie2019}
N.~Lagaillardie, M.~A. Djari, and O.~G\"{u}rcan, ``A computational study on
  fairness ofthe tendermint blockchain protocol,'' {\em Information}, vol.~10,
  no.~12, 2019.

\end{thebibliography}

\end{document}